# *P* values, confidence intervals, or confidence levels for hypotheses?

11 February 2014

Michael Wood
University of Portsmouth Business School
Richmond Building
Portland Street, Portsmouth
PO1 3DE, UK
michael.wood@port.ac.uk
mickofemsworth@gmail.com

## Abstract

Null hypothesis significance tests and *p* values are widely used despite very strong arguments against their use in many contexts. Confidence intervals are often recommended as an alternative, but these do not achieve the objective of assessing the credibility of a hypothesis, and the distinction between confidence and probability is an unnecessary confusion. This paper proposes a more straightforward (probabilistic) definition of confidence, and suggests how the idea can be applied to whatever hypotheses are of interest to researchers. The relative merits of the different approaches are discussed using a series of illustrative examples: usually confidence based approaches seem more transparent and useful, but there are some contexts in which *p* values may be appropriate. I also suggest some methods for converting results from one format to another. (The attractiveness of the idea of confidence is demonstrated by the widespread persistence of the completely incorrect idea that *p*=5% is equivalent to 95% confidence in the alternative hypothesis. In this paper I show how *p* values can be used to derive meaningful confidence statements, and the assumptions underlying the derivation.)

# *P* values, confidence intervals, or confidence levels for hypotheses?

## Introduction

Null hypothesis significance tests (NHSTs) are widely used to answer the question of whether empirical results based on a sample are due to chance, or whether they are likely to indicate a real effect which applies to the whole population from which the sample is taken. There are, however, serious difficulties with such tests and their resulting *p* values or significance levels: the literature on these difficulties goes back at least half a century (e.g. Cohen, 1994; Gardner and Altman, 1986; Gill, 1999; Kirk, 1996; Lindsay, 1995; Mingers, 2006; Morrison and Henkel, 1970; Nickerson, 2000). NHSTs are very widely misinterpreted, they do not provide the information that is likely to be wanted, and as many null hypotheses are obviously false the tests are often unnecessary as well as uninformative.

The commonly suggested alternative to the use of NHSTs is the use of confidence intervals (e.g. Cashen and Geiger, 2004; Cortina and Folger, 1998; Gardner and Altman, 1986; Gill,1999; Mingers, 2006; Wood, 2005). In medicine, for example, guidance to authors of research papers in some journals (BMJ, 2011), and regulatory authorities (ICH, 1998), strongly recommends these in preference to NHSTs. However, in most of the social sciences, NHSTs, and not confidence intervals, are still the standard.

There are, however, also problems with confidence intervals:

1    They refer to an interval whereas in many cases researchers do want to evaluate a hypothesis.

2    "Confidence" is usually defined in a rather awkward way which appears to distinguish the concept from the probabilities that people intuitively want.

3    They are inapplicable if the characteristic of interest cannot be expressed on a suitable numerical scale.

My aim in this paper is to extend the idea of confidence to include confidence levels for hypotheses in general (not just intervals), to propose that confidence levels can reasonably be interpreted as probabilities, to suggest some simple methods for deriving confidence levels from *p* values, and to assess the relative merits of NHSTs, confidence intervals and confidence levels for hypotheses. This should be of interest to any researcher concerned about the best way to analyze and communicate statistical results.

For example, according to a study which sought to investigate the impact of social status on mortality by analyzing how winning an Academy award (Oscar) may prolong an actor's life,

"life expectancy was 3.9 years longer for Academy Award winners that for other, less recognized performers (79.7 vs. 75.8 years, $p = 0.003$)" (Redelmeier and Singh, 2001: 955). The *p* value here does not directly address the question of how likely it is that Oscar winners really do live longer – the equivalent confidence level for this hypothesis is 99.85% (making a few reasonable assumptions to be explained below). Alternatively, we might cite a confidence level for the slightly stronger hypothesis that the life expectancy of Oscar winners is at least a year longer (98.6%). Confidence levels of this type avoid most of the difficulties of *p* values – they do, for example, seem far easier to understand.

I start with a brief discussion of the concepts used to frame the problem that all the methods are tackling – that of using a sample to make inferences about a wider population. I then review briefly the difficulties of NHSTs, how confidence intervals overcome many of these difficulties and how confidence can be defined. Then I explore the idea of confidence levels for more general hypotheses and how they can be estimated. I finish with a discussion of a series of examples, chosen to illustrate the advantages and disadvantages of *p* values, confidence intervals, and confidence levels for hypotheses in a range of different contexts. My concern in this article is with the concepts used to express statistical conclusions, not with the detail of methods of analysis; I have chosen examples using relatively simple methods because this makes it easier to analyze these concepts.

## Samples, populations, processes and the wider context

Hand (2009: 291) points out that "much statistical theory is based on the notion that the data have been randomly drawn from a population" but this is often not the case. To make sense of statistical inference procedures we then need to imagine a population from which the sample can reasonably be assumed to have been randomly selected. The sample of Oscar winners included only past Oscar winners, but the aim of the research was to see if anything could be inferred about the life expectancy of Oscar winners in general, including future winners. We then need to make the assumption that the sample can be regarded as a random sample from this population of current and potential future winners – which is obviously a difficult notion to nail down precisely.

Experiments, or randomized trials, also make the idea of the population problematic. Suppose, for example, we are comparing two training programs, A and B, with the help of 42 trainees[1]: as I will use this as an example below it is helpful to give a few details. A randomly chosen group of 21 trainees does Program A, the remainder to do B, and then the effectiveness of the training for each trainee is rated on a 1-7 scale. To compare the two programs we then work out the mean effectiveness ratings for each group of 21 trainees: these come to 4.48 for Program A and 5.55 for Program B. This suggests that Program B is better, but does not answer the question of whether the effect may be due to chance and might be reversed in another similar experiment. In one sense the population here is the wider group of trainees from whom the sample is drawn, but even if the 42 trainees comprised the entire population, the problem of sampling error still arises because a different division of trainees into two groups might produce a different result.

In both cases it is obviously important to analyze how reliable the result is taking account of sampling error, although the idea of a population is difficult to visualize. An alternative metaphor is the idea of a "process": we could refer to the training process or the process of winning an Oscar. This is still rather awkward, and does not acknowledge the future dimension. A term such as "wider context" is vaguer, and so more suitable for informal descriptions, although for formal work the notion of population is convenient and deeply embedded in the language of statistics. For the Oscars, the wider context includes the future, and for the training programs the wider context includes both a wider population of trainees and the fact that there are many possible allocations into two groups.

## Null hypothesis significance tests (NHSTs) and p values

The idea of an NHST is to set up a null hypothesis and then use probability theory or simulation to estimate the probability of obtaining the observed results, or more extreme results, if the null hypothesis is true[2]. If this probability, known as the *p* value, is low, then we conclude the null hypothesis is unlikely and an alternative hypothesis must be true. Many researchers use cut-off levels of 5%, 1%, etc and describe their results as significant at 5% (p<0.05) or whatever. The result above about Oscars and life expectancy, for example, is significant at 1%, indicating reasonably strong evidence against the null hypotheses, and so for the hypotheses that winning an Oscar really does tend to prolong life. On the other hand, a subsequent analysis using a difference method of analysis and including more recent data gave results that are equivalent to a *p* value between about 13% and 17% (Sylvestre et al, 2006; I have estimated the *p* value from the confidence interval given in the article, using the method described below). This suggests that the difference in life expectancies is well within the range that would be expected if only chance factors were at work. In contrast to the earlier result, this suggests there is little evidence for the hypothesis that winning an Oscar prolongs an actor's life.

Similarly, the advantage of program B over Program A yields a *p* value of 2.11%. This is shown graphically in Figure 1[3]. This graph represents the likely distribution of the mean difference between the programs in similar samples drawn from the same source, *assuming* the truth of the null hypothesis that the difference between the population means is actually zero, and that any difference observed in the sample is just due to random sampling error (this is often called a sampling distribution). The graph shows that the probability of this difference being as big as, or bigger than, the observed difference of 1.07 is only 2.11%. As this *p* value is low we can assume the null hypothesis is unlikely and there is a real difference in the effectiveness of the two programs.

**Figure 1: Probability distribution for sample estimate of difference between Program B and Program A *assuming* the null hypothesis of no population difference**

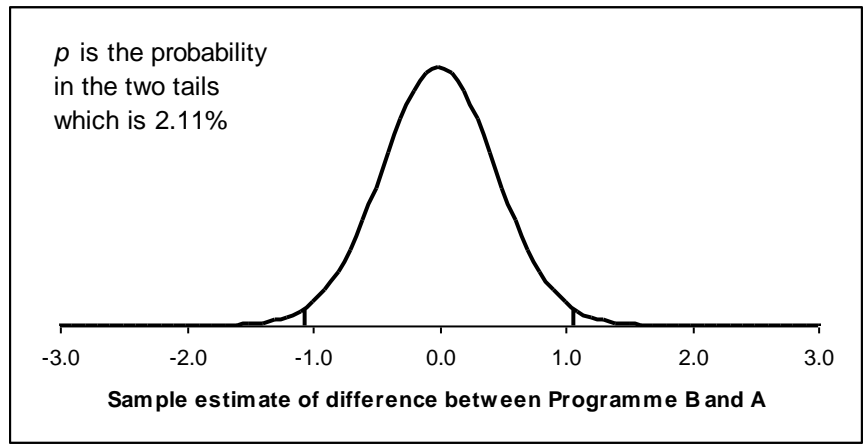

*p* is the probability in the two tails which is 2.11%

**Sample estimate of difference between Programme B and A**

As noted above, NHSTs have attracted some very extensive criticism over the years. I will review some of the main points here, but for a more extensive review the reader is referred to the citations above. There are counter-arguments; taken in the right spirit, sometimes NHSTs may have a useful role to play – I will discuss this in relation to some specific examples below.

1. *NHSTs fail to provide the information people are likely to want.* The *p* value from the NHST about Oscar winners does not tell us (a) how many extra years Oscar winners are likely to have (the strength of the effect), and (b) how probable it is that winning an Oscar winner does increase life expectancy (the *p* value may give us some indication but it does *not* give this probability). The *p* value concerning the two training programs is similarly uninformative. Readers of the article about the two training programs are given the mean scores for each group, but the difference between the means is not made explicit in the article. This is part of a general tendency for "quantitative" research in some social sciences to be strangely non-quantitative in the sense that readers are often not told the *size* of effects and differences. This is not, however, true of the article about the Oscars (in a medical journal with different conventions) where the 3.9 years is made explicit.

2. *The conclusions that can be drawn from NHSTs are often trivial.* Strictly, the null hypotheses on which *p* values are based are exact: the Oscar winners' life expectancy and that of the controls are exactly the same, and the population mean scores for both training programs are identical. In practice, slight differences are likely between any two groups, so null hypotheses of this kind are very likely to be false, which means that there is little point in a formal test to prove it. The result will depend on the sample size: with a suitably large sample almost any null hypothesis is likely to be disproved. Even apart from this logical point, null hypotheses are sometimes so unlikely as to make disproving them of marginal interest. For example, Grinyer et al (1994) tested the – very implausible – hypothesis that respondents to a questionnaire are equally likely to agree with a statement, or disagree, or neither agree nor disagree; and Glebbeek and Bax (2004: Table 2) cite a *p* value less than 1% for the relationship between employee turnover and the performance of organizations – common sense and much of the literature suggests that a null hypothesis of no relationship between these two variables is false, so the *p* value adds little.

3. *NHSTs are very widely misinterpreted.* Statistically significant results are widely assumed to be large and important in a practical sense. The fact that a *p* value is 5% is widely viewed as implying that the probability of the truth of the alternative hypothesis is 95%. A non-significant *p* value is often seen as some sort of proof for the truth of the null hypothesis (this fallacy is built into the "test of normality" often used as a pre-check for some statistical procedures). None of these are valid. Nickerson (2000) lists ten distinct ways in which NHSTs can be, and often are, misunderstood. Part of the reason for this is doubtless the natural tendency to assume that a carefully crafted statistic like a *p* value will deliver the information that is obviously wanted: unfortunately this is not the case. Coulson et al (2010) tested how well 330 authors of published articles understood *p* values and confidence intervals. They concluded that "interpretation was generally poor". However, there was very clear evidence that many authors interpreted confidence intervals in terms of *p* values; those who interpreted confidence intervals without reference to null hypothesis tests gave a far better interpretation of the results than those who thought in terms of null hypothesis tests, which suggests that NHSTs are a powerful confusing influence in the interpretation of statistics, even among professional researchers.

    The terminology commonly used is not helpful. "Significant" in ordinary English does mean important. Widely used phrases like A "is significantly more than" B suggests that the statistical significance is a property of the difference between A and B, as opposed to being just a measure of the strength of the evidence for this difference. Furthermore, *p* values mean focusing on a hypothetical null hypothesis instead of the hypothesis of interest. And to cap it all, as a measure of the strength of evidence, *p* values are a reverse measure – *low* values indicating *strong* evidence. All these factors make the widespread misunderstanding of NHSTs seem almost inevitable.

# Confidence intervals

The idea of confidence intervals is to use the data to derive an interval within which we have a specified level of confidence that the population parameter will lie. For the Oscars, the first analysis suggests that the extra life expectancy for winners is 3.9 years and the 95% confidence interval for this additional life expectancy is likely to be about 1.3 to 6.5 years (an estimate from the results given by Redelmeier and Singh (2001) using the normal distribution as described below). We cannot be sure of the exact advantage from winning an Oscar on the basis of the sample data, but we can be 95% confident that the true figure will lie in this interval.

I will look at the second example in more detail and carry this through to the discussion of confidence levels. Figure 2 shows a *confidence distribution* for the population difference between the means of the effectiveness ratings of Programs A and B. This is derived from Figure 1 simply by shifting the curve along so that it is centered on 1.07 (the observed difference of the means) rather than 0. An informal rationale for this goes as follows. The most likely value for the population parameter, given the sample data, is the sample estimate (1.07), so it makes sense that this should be the centre of the confidence distribution. Furthermore, Figure 1 suggests that the probability that the difference between the sample estimate, and the unknown population value (assumed to be 0), being more than 1.07 is 2.11%, so the tails in Figure 2 are correct from this perspective. Figure 1 can be regarded as describing the probabilities of different discrepancies between the sample estimate and the population parameter being estimated, so it is reasonable to regard a displaced version of Figure 1, in Figure 2, as representing our view, based on the sample information, of different possible values of the population parameter. The horizontal axis in Figure 2 refers to the possible values of the unknown population parameter (the difference of the means), whereas in Figure 1 it refers to sample estimates. Figure 2 (and the spreadsheet behind it) enables us to read off the 2.5 and 97.5 percentiles of this distribution – this is the 95% confidence interval which extends from 0.17 to 1.97. (There is a more detailed discussion of the rationale behind this in the Appendix.)

**Figure 2: Confidence distribution and interval for difference between Program B and Program A**

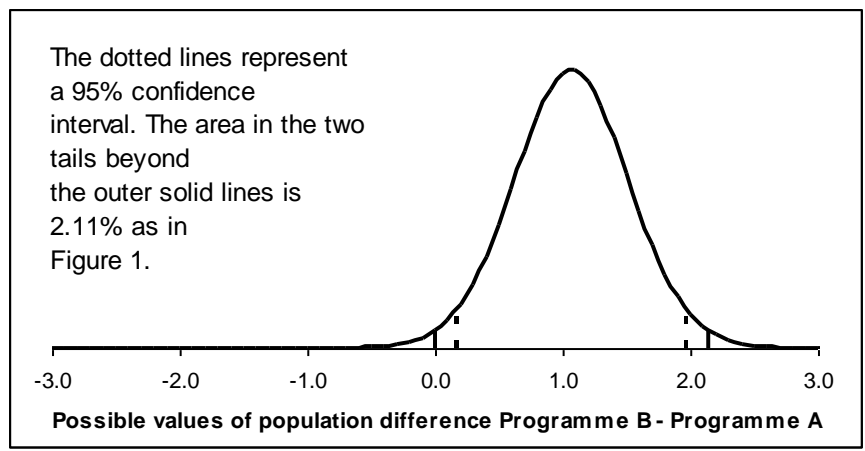

The dotted lines represent a 95% confidence interval. The area in the two tails beyond the outer solid lines is 2.11% as in Figure 1.

**Possible values of population difference Programme B - Programme A**

This has clear advantages over the *p* value presentation. It does answer directly questions about the strength of the effect (how big the difference is), and the width of the interval describes the uncertainty due to sampling error in an obvious way. The information displayed is not trivial or obvious like the NHST conclusions may be, and misinterpretations seem far less likely than for NHSTs. The focus is very much on the difference between the programs and not on a hypothetical null hypothesis, there is no inverse scale, and the phrase "confidence" suggests that what is being assessed is the strength of the evidence.

It is also worth noting that, because zero is not in this interval, we can be more than 95% confident that Program B is better than Program A, which is equivalent to the statement that p<0.05. In this way all the information in the significance level can be deduced from confidence intervals, but the confidence intervals provide extra information about the size of the difference and the extent of the uncertainty.

Confidence intervals corresponding to many other null hypothesis models have been derived and built into software packages. They are widely used as a means of analyzing and presenting results in some fields such as medicine, but not in most social sciences.

## Confidence levels for hypotheses

The notion of confidence can easily be extended from intervals to more general hypotheses. This idea can easily be applied to the difference between Programs A and B (Figures 1 and 2). The confidence curve in Figure 2 is symmetrical so the confidence of the value being in the lower tail will be half of 2.11% or 1.1% (rounded to one decimal place), and the confidence that the difference in the means will be greater than zero, or that Program B is actually, in population terms, better than Program A is 100% − 1.1% = 98.9%.

This principle can easily be extended. Suppose we were interested in the hypothesis that the advantage of Program B is substantial, say greater than one unit. Then the *p* value is no longer helpful, but in the case of Figure 2, we can use the *t* distribution (more details below) directly to show that:

Confidence (Program B more than 1 unit better than A) = 56%.

It should be clear from Figure 2 that this is roughly right.

In a very similar way, using Redelmeier and Singh's (2001) data and methods, the confidence level for the hypothesis that Oscar winners live longer than the controls is 99.85%, and the confidence for their life expectancy being at least a year longer is 98.6%.

The idea of a confidence level for a hypothesis is more general than these two examples might suggest. For example, Glebbeek and Bax (2004) wanted to confirm the hypothesis that there is an "inverted U-shape relationship" between two variables – staff turnover and organizational performance – by setting up regression models with both staff turnover, and staff turnover squared, as independent variables. Because this hypothesis does not depend on a single parameter, it is awkward to use *p* values or confidence intervals to support this hypothesis. Glebbeek and Bax (2004) actually used *p* values, but it is easy to use a bootstrap argument to estimate the confidence level for this hypothesis – this comes to 67% (Wood, 2012).

## Confidence as probability

The word "confidence" is conventionally used to indicate that the concept is *not* probability but is to be interpreted in frequentist terms. This means that we need to imagine repeating the procedure that led to a 95% confidence interval, for example: then if the 95% is accurate, 95% of these repetitions should produce an interval which includes the true value of what we are trying to estimate (see, for example, Bayarri and Berger, 2004). On the other hand, interpreting a 95% confidence interval as a probability would simply involve asserting that there is a probability of 95% that the truth about the whole population lies somewhere in this interval. This distinction is described as "subtle" by Nickerson (2000, p. 279), and is one of the issues at stake in the literature on the foundations of statistical inference.

This literature is complex both conceptually and mathematically, and has spawned debates without easy answers. One influential and important perspective is the Bayesian one: the Bayesian equivalent of a confidence interval is a credible interval. These are sometimes identical to frequentist confidence intervals (Bayarri and Berger, 2004: 63), so in these cases it would be reasonable to view confidence levels as probabilities, and to identify the confidence distribution with the posterior probability distribution. The confidence level for a hypothesis is simply the posterior probability of the hypothesis.

Part of the reason for the reluctance to do this stems from the frequentist view that either the population mean is in the confidence interval, or it is not, and that the idea of probability cannot meaningfully be used to express this type of uncertainty. However, in everyday discourse the idea of using probability for this type of "epistemic" uncertainty is widespread and unproblematic, so it would seem sensible to ignore the frequentists' philosophical objections and treat confidence levels as probabilities.

Another reason for the reluctance to use Bayesian methods is that these bring in prior probabilities to reflect prior beliefs about the situation. In practice, this may be difficult, and is widely seen as introducing an unhelpful element of subjectivity into the calculations. However, the assumption necessary to produce Bayesian intervals which are identical to standard confidence intervals for the example in Figure 2 is that the prior probabilities should be uniform indicating that all values on the horizontal axis in Figure 2 are equally likely (see Appendix). In Bayesian terms, this is the assumption on which the derivation of Figure 2 and the above confidence interval depends.

My suggestion is that we adopt the following *definition* of a confidence level for an interval or hypothesis:

A confidence level is defined as an estimate of the *probability* of the true value of the parameter being within the interval, or of the *probability* of the truth of the hypothesis.

There may be different ways of estimating this probability: using Bayesian credible intervals based on a uniform prior distribution, or on some other prior distribution, or by the methods used to derive confidence intervals. Obviously, different methods of computation may give slightly different answers, but this is hardly unusual in statistics where many concepts are slippery and can be made precise in different ways. (Different statistical tests may give different *p* values, for example.) The Bayesian methods have the advantage that they are arguably more transparent: Bayarri and Berger (2004) suggest that the Bayesian approach should be "taught to the masses" (p. 59) and that this is often possible "without changing the procedures that are taught".

## Methods of estimating confidence levels

If we think of a confidence distribution as a Bayesian posterior distribution, then we have the whole gamut of Bayesian methods at our disposal. Similarly there are a wide variety of methods for deriving confidence intervals, which could easily be adapted to give confidence levels for more general hypotheses. This includes bootstrapping, an approach of very general applicability, which was used to generate the 67% confidence level mentioned above.

However, in practice, we may be using software which just generates *p* values, and possibly confidence intervals. Alternatively we may just have the results in a published paper. The methods outlined below are for estimating confidence levels from the information we are likely to have in these circumstances.

### Estimating confidence levels from *p* values or other NHST statistics

If we only have *p* values (from a software package or a published paper) it is sometimes possible to estimate confidence levels as we have seen above. This approach assumes that it is reasonable to shift the null hypothesis distribution and treat it as a confidence distribution as described above in the section on confidence intervals. There is a more detailed analysis of the conditions under which this is reasonable in the Appendix: in rough terms the *curve shift method* is likely to be reasonable if the null hypothesis is modeled by a symmetrical distribution such as the normal or *t* distributions.

The method above can easily be generalized. If the difference of the means were negative, the argument above will be reversed, so in general we can write

Confidence (Pop. parameter > 0)     $= 1 - p/2$     *if* Sample estimate ≥0

$= p/2$     *if* Sample estimate <0

where *p* is the two tailed *p* value for the null hypothesis that the population parameter is zero.

In Figure 1 the *one-tailed p* value is 2.1%/2 = 1.1% which is the same as the confidence that Program A is better than Program B. Confidence levels thus give us another way of interpreting one-tailed significance levels. On the other hand, the argument here is *not* consistent with the common misconception that *p* is 5% means that the probability of the null hypothesis being true is 5%, so the probability that the probability that the alternative hypothesis is true – that there is a difference between programs A and B – is 95%. Figure 2 illustrates the problem with this. The probability of the difference between the two programs being *exactly* zero is very small indeed, certainly not 5%.

In some cases *p* values are not given exactly but as an inequality. The example above mirrors Eggins et al (2008), where the *p* is given a single star, indicating that *p* < 5%. This means that the above reduces to the assertion that the confidence level is greater than 97.5%.

If we are starting from a value of *t* or *z*, or if we want a confidence level for another hypothesis, it is possible to deduce the standard error from the given information and then use this to calculate confidence levels from the *t* or normal distributions. The arithmetic here is easy, and is incorporated in the spreadsheet at http://woodm.myweb.port.ac.uk/CLIP.xls.

### Estimating confidence levels from confidence intervals

Sylvestre et al (2006), in their paper on Oscar winners, gave a 95% confidence interval for additional life expectancy enjoyed by Oscar winners as −0.3 years to +1.6 years. The confidence level for the hypothesis that this additional life expectancy is positive could be estimated by rerunning the analysis with different confidence levels until one is found that has a lower limit of zero, which can then be used to estimate the confidence level for the hypothesis that the additional life expectancy is positive and that the experience of winning an Oscar is linked to longer life expectancy. In this way any software package generating different confidence intervals could be used to build up a confidence distribution.

In practice, it may be reasonable to assume that the confidence distribution is approximated by a *t* or normal distribution, in which case the mean and one of the given limits can be used to estimate the standard error and hence use the *t* or normal distribution to estimate any confidence level we want – the confidence level for the hypothesis that the additional life expectancy is positive is somewhere between 91% and 94% (using the spreadsheet at http://woodm.myweb.port.ac.uk/CLIP.xls, the answer depending on which limit we take as they are not quite symmetrical). This is obviously just an approximation, but there seems little point in being too pedantic when the interpretation of confidence is likely to refer to arbitrary levels such as 95%, and estimates of statistics such as *p* values are themselves subject to considerable variation between samples (Boos and Stefanski, 2011).

## Some examples

I will start by drawing together the discussion of the examples above, and then consider a few more examples – chosen to illustrate a number of issues. All statistics not explained below, or given by the authors of the original research, are estimated roughly using the methods described in the previous section and the spreadsheet at http://woodm.myweb.port.ac.uk/CLIP.xls.

## Oscars and life expectancy

Redelmeier and Singh (2000) found that Oscar winners' life expectancy was 3.9 years longer than the controls. They cited a *p* value for this result:

> *p* = 0.003

Alternatively they could have stated that:

> 95% confidence interval for additional life expectancy is 1.3 to 6.5 years
> Confidence level for positive additional life expectancy = 99.85%
> Confidence level for additional life expectancy of one year or more = 98.6%

The equivalent results for Sylvestre et al's (2006) updated analysis are[4]:

> *p* = 0.15
> 95% confidence interval is -0.3 to 1.7 years
> Confidence level for positive additional life expectancy = 93%
> Confidence level for additional life expectancy of one year or more = 27%

The confidence intervals and levels seem more useful and easier to interpret than the *p* values.

## Training programs A and B

This example was introduced because it is an experiment, or randomized controlled trial. However, the issues regarding the analysis are similar to the Oscars example and to the next example, so I will not analyze this further here.

## Men and women in work-home culture

This example is included to show how the results in a typical article in a social science journal, the *British Journal of Management,* could be analyzed in terms of confidence. As part of a study of "work-home culture and employee well-being" Beauregard (2011) showed the differences between men and women on 9 variables in her sample of 224 local government employees. I will use two of these variables as illustrations:

**Table 1. Some of the results in Table 1 in Beauregard (2011)**

| Measure | Mean for men (n=84) | Mean for women (n=140) | t(222) |
|---|---|---|---|
| **Work-home culture: managerial support** | 4.34 | 4.56 | − 1.33 |
| **Hours worked weekly** | 41.27 | 36.69 | 3.68*** |

She also gives the SD for each variable, and a note under the table explains that *** means p<0.001: the difference between men and women is not significant (p>0.05) for the first variable, and highly significant for the second. As is usual in management research, confidence intervals are not given.

Tables 2 and 3 show the same results in terms of confidence intervals and levels.

**Table 2. Some of the results in Table 1 in Beauregard (2011) expressed as confidence intervals**

| Measure | Mean for men (n=84) | Mean for women (n=140) | Difference of means (Men – Women) | 95% Confidence Interval for difference of means |
|---|---|---|---|---|
| Work-home culture: managerial support | 4.34 | 4.56 | -0.22 | -0.55 to +0.11 |
| Hours worked weekly | 41.27 | 36.69 | 4.58 | 2.1 to 7.0 |

**Table 3. Some of the results in Table 1 in Beauregard (2011) expressed as confidence levels for hypotheses**

| Measure | Mean for men (n=84) | Mean for women (n=140) | Difference of means (Men – Women) | Confidence level for the hypothesis: Mean for Men > for Women |
|---|---|---|---|---|
| Work-home culture: managerial support | 4.34 | 4.56 | -0.22 | 9% |
| Hours worked weekly | 41.27 | 36.69 | 4.58 | 99.99% |

**Telepathy**

In a series of experiments in the 1920s and 30s, the psychologist, J B Rhine, found a number of people who appeared to be telepathic (Rhine, 1997). In one series of experiments, Hubert Pearce Jr. did a card guessing experiment 8,075 times, and got the card right on 3,049 occasions. There were five cards in the pack, so guesswork would have produced about 1615 hits. Rhine argues that Pearce's performance is so much better than guesswork that telepathy must be involved; others have taken the hypothesis that Pearce was cheating more seriously (Hansel, 1966). We can model the number of correct cards under the null hypothesis that Pearce was guessing using the binomial distribution, and then use the normal approximation ($z$ = 39.9) to deduce that the two tail $p$ value is, for all practical purposes, zero, which means that the results could not have arisen from chance alone.

This is an example where the $p$ value does seem entirely appropriate and the idea of confidence would be rather awkward: the reasons for this are discussed below.

**Heart transplants**

In October 2007, heart transplants were stopped at Papworth Hospital in the UK because 7 out of 20 patients had died within 30 days of their operation (Garfield, 2008): this was significantly more than the national average rate of about 10% ($p$ = 0.0024, one tail, using the binomial distribution with a mean of 2 deaths in a 20 patient group to model the null hypothesis as shown in Figure 3[5]).

**Figure 3. Distribution of number of deaths in 20 patient groups assuming the null hypothesis that the mean death rate = 10%**

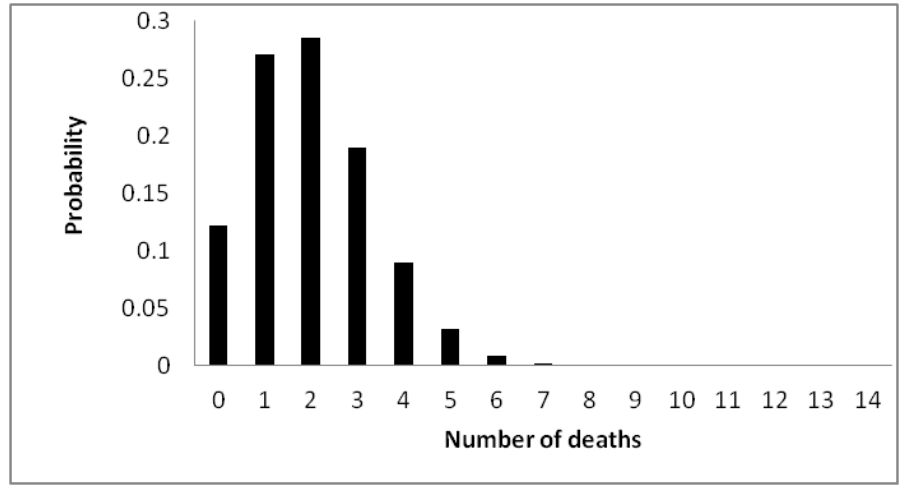

The curve shift method obviously cannot work here. If we shift the curve five units to the right so that it is centered on the sample value of 7, this clearly cannot represent a confidence distribution because it would indicate that there is zero confidence that the population mean is 0, 1, 2, 3 or 4 deaths, whereas 4 deaths in particular does seem reasonably consistent with the sample value of 7. The binomial distribution is a different shape for different population means, so we cannot simply slide it along.

Confidence intervals and levels here clearly need to be estimated using other methods: the standard normal approximation method gives a 95% confidence interval (based on 7 out of 20, or 35%, of patients dying) extending from 14% to 56%, a numerical Bayesian approach (using the spreadsheet at http://woodm.myweb.port.ac.uk/ConfIntsPoissonBinom.xls) gives an interval extending from 18% to 57%, and there are a number of other methods giving similar answers.

These same methods can be used to estimate a confidence level for the hypothesis that the long term death rate at Papworth Hospital is more than 10%: this comes to 99.8% by the first method above, and 99.94% by the second.

**Staff turnover and organizational performance**

We have mentioned above the estimated confidence level of 67% for the hypothesis that this relationship has an inverted U-shape with very high and very low levels of staff turnover both leading to suboptimal performance (Glebbeek and Bax, 2004; Wood, 2012). There is no satisfactory and easy way of using *p* values or confidence intervals to express this result.

Glebbeek and Bax (2004) also tried a linear (straight line) model of the relationship between staff turnover and performance (with three control variables): the regression coefficient in one model was -1778: this means that predicted performance fell by 1778 currency units per staff member for each additional 1% in staff turnover. The *p* value found was 0.007, which can be converted to a 95% confidence interval extending from -3060 to -495 currency units per additional 1% of staff turnover, or to a confidence level that the regression coefficient is positive of 99.65%.

## Discussion: a comparison of p values, confidence intervals and confidence levels for hypotheses

Most of the above examples can be analyzed by all three approaches. There are two examples where this is not so. For the inverted U-shape hypothesis relating staff turnover and organizational performance there is no easy and obvious way to use *p* values and confidence intervals, so the confidence level approach is the obvious one to use.

With the telepathy example, to use the data to derive a confidence interval or level we need to define a suitable measure to assess the extent to which telepathy is occurring. The obvious statistic is the proportion of correct guesses: under the guessing hypothesis we would expect population value of this measure to be *exactly* 20%, and under the telepathy hypothesis it would be more than 20%. If we were going to use confidence intervals, 20% would be just one value on the continuum of possibilities, which would mean that the confidence level for the hypothesis that the proportion of correct guesses is *exactly* 20% is zero – which is not a helpful answer. This effectively rules out the idea of confidence as discussed in this paper. Furthermore, this is one instance in which *p* values are not trivial and do make good logical sense, because the null hypothesis is an *exact* one, and *any* departures from 20% are surprising from this point of view. This suggests the *p* value as the appropriate statistic here.

This example illustrates neatly the main advantage of NHSTs over their rivals: the underlying rationale is straightforward involving the estimation of a probability under the assumptions of the null hypothesis. There are none of the extraneous, and possibly questionable, additional assumptions which are necessary to use the idea of confidence – foremost among these is the assumption, explicit in the Bayesian formulation and implicit in frequentist formulations, that all possibilities are assumed to be equally likely before analyzing the evidence. NHSTs and *p* values may not be user-friendly and may just tell a small part of the story, but for thoughtful users, the rationale is simpler and involves fewer assumptions.

Let's now consider Men and women in work-home culture (Table 1). The confidence interval presentation (Table 2) has the advantage of telling readers how strong the effect is, and the likely level of uncertainty due to sampling error. The *p* values given in the original article do not directly tell readers how big the difference is, nor the likely impact of sampling errors on the result, and they are difficult to interpret for the reasons discussed above. The comparison with confidence levels (Table 3) is less clear cut because simply telling readers that there is 99.99% chance that men (in this population) work longer hours than women says nothing about the size of the effect – how much longer they work. The confidence interval presentation here is arguably the most informative, with the confidence level presentation providing a simple summary in terms of the hypotheses of interest to the researchers.

Very similar arguments apply to the difference between the two training programs, the heart transplants, the linear model relating staff turnover and organizational performance, and the Oscars and life expectancy. In the last of these examples, I have shown how confidence levels for hypotheses can be made more useful by considering different hypotheses. The confidence levels for Oscar winners living at least a year longer tell a slightly different story from the confidence levels in them simply living longer. Another possibility, of course, would be to show a graph of the confidence distribution (like Figure 2).

One important way in which the examples vary is in terms of the *status of the null hypothesis.* For many people telepathy is so unlikely that the alternative, null or chance hypothesis, is very much the front runner. The situation with the heart transplants is rather different in that there are fairly obvious reasons for differences between hospitals, but it still makes excellent sense to take the national average as a baseline for comparison. In both cases there are good reasons for taking the null hypothesis seriously, and so, from this perspective, NHSTs are a reasonable approach (although in the latter case the arguments against them may be stronger).

This is not true of the other examples. There is very little reason to think that staff turnover would have no impact on performance, or that there would be no difference between men and women in work-home culture, or that two training programs would be (exactly) equally effective. These null hypotheses have little interest or credibility, which, means, firstly that testing them is of marginal interest, and secondly that the focus on the null hypothesis is likely to seem odd to readers of the research (this is possibly acknowledged by the common practice of not mentioning null hypotheses in research publications). In these three examples, it definitely makes sense to focus on confidence, because then the focus is on the hypothesis or interval of interest; there is no strange, hypothetical and distinctly uninteresting null hypothesis involved.

## Conclusions

Using confidence intervals, or giving confidence levels for hypotheses of interest, has the potential to avoid many of the widely acknowledged problems of NHSTs and *p* values. Confidence seems a more intuitive and direct concept which avoids the need to formulate a null hypothesis in order to demonstrate how implausible it is.

For example, research studying the hypothesis that Oscar winners live longer is better served by giving a confidence level for this hypothesis (98.5% for one study, 93% for the second) than a *p* value because the former gives a direct estimate of the probability of the hypothesis being true whereas the latter does not. To convey information about the size of the effect, we could either use a confidence interval, or give the confidence level for Oscar winners having at least one year additional life expectancy (98.6% for the first study and 27% for the second).

In a more typical social science context, instead of the conventional *t* and *p* values in tables such as Table 1, we could use confidence intervals as in Table 2, or confidence levels as in Table 3. Using confidence intervals or levels involves converting the characteristic of interest to a single quantity – typically a difference between two means, or a regression coefficient (slope).

The idea of confidence is conventionally viewed as distinct from the idea of probability, but I have argued above that this is unnecessary. Confidence distributions could be *defined* as probability distributions: the probabilities in question could then be estimated in a variety of different ways, including as Bayesian posteriors based on a flat prior distribution. This may offend purist statisticians, but it is worth remembering that the typical underlying assumption of a random sample from a large population may bear only a very rough relationship with reality, and that empirical estimates of both confidence levels and *p* values are themselves uncertain and unreliable estimates.

In practice, because of the current dominance of null hypothesis testing, the information we often have comprises *p* values and statistics such as *t* and *z*. Under many circumstances (see the Appendix) it is reasonable to estimate confidence distributions, and so confidence intervals and levels for other hypotheses, by shifting the null hypothesis distribution along so that it represents a confidence distribution – e.g. for the difference of two means or proportions, for regression coefficients, or any other statistic for which the *t* or *z* distribution is the basis of the null hypothesis test. In these cases there is a very simple formula for deriving confidence levels for the hypothesis that the population value of the statistic is above or below zero (or other null hypothesis value) – either $p/2$ or $1 - p/2$. For other hypotheses and confidence intervals, it is be possible to use a given *p* value to "reverse engineer" the confidence distributions and so derive the required statistics – a spreadsheet is available (http://woodm.myweb.port.ac.uk/CLIP.xls) for performing the simple calculations involved.

Despite these arguments, NHSTs do have certain advantages. *P* values are probabilities of certain events happening on the assumption that the null hypothesis is true; in terms of detailed rationale this is conceptually simpler than confidence based methods because these depend on an argument involving various assumptions (like the flat priors assumption) to derive confidence from probability. If the null hypothesis is a credible or interesting hypothesis, then *p* values do make some sense. However, for a non-technical audience (i.e. almost everybody) it would be sensible to avoid jargon like "*p*" or "significant" and use phrases like "the probability of getting a sample value this far from 0 is 0.3% if only chance factors are at work", or "the data is consistent with the hypothesis that there is no difference and only chance factors are at work." And, of course, readers also need to know how big the difference, or other measure of effect, is. This may necessitate a lengthier description of conclusions, but hopefully one that is more informative and less likely to lead to misunderstanding. If we want a probability for the truth of our hypothesis, then we need a confidence level, not a *p* value. In practice, research results could easily be given in several formats, which may be the best way of comparing the practical value of the different approaches.

## Appendix: A Bayesian analysis of the validity of the curve shift method

Let's suppose that we are interested in a numerical population parameter, $\theta$, and we have some sample information which gives an estimate of its value – say $\hat{\theta}$. The typical null hypothesis would be that the population value of $\theta$ is zero, but $\hat{\theta}$ is typically slightly different from zero. Regardless of whether the range of possible values for $\theta$ is discrete or continuous, we can regard it as discrete if we remember that there is a limit to the accuracy of our measurements. Furthermore, in practice, there is likely to be a minimum possible value and a maximum. This means that we have a finite number, $n$, of possible hypotheses about the value of $\theta$ which we can call $H_{min}$ … $H_{max}$. For example, in Figure 1, if we decide to measure to the nearest tenth, and assume that -10.0 is the minimum possible value of $\theta$, and +10.0 is the maximum, then, for example, the 101st hypothesis, $H_{0.0}$, is the null hypothesis that $\theta$ is zero, and the 102nd hypothesis is $H_{0.1}$, the hypothesis that $\theta$ is actually 0.1. Similarly, if we imagine taking further samples, there are $n$ possible sample estimates of $\theta$; the estimate from the actual sample, $\hat{\theta}$, being 1.1.

We now make the following assumptions:

1. There is a distribution curve for the sample estimates, under the null hypothesis, which extends far enough for the curve to be shifted. If, for example, the parameter being estimated were a correlation coefficient, the shift might move parts of the curve above +1 or below -1, which, of course, cannot be correct because correlation coefficients cannot take values outside these limits. Also, there has to be the possibility of departures from the null hypothesis in both directions (otherwise the probability of departures in the possible direction will simply be 100%) – this rules out the $\chi^2$ test for goodness of fit, for example.
2. This distribution is symmetrical.
3. The probability distributions for the sample estimates under each of the hypotheses, $H_i$, are the same shape and width as the distribution under the null hypothesis.
4. The prior probabilities for all hypotheses, $H_i$, are equal.

   Bayes theorem now tells us that the posterior probability of hypothesis $H_i$ is

   $$P\left(H_i \mid X = \hat{\theta}\right) = \frac{P(X=\hat{\theta}|H_i)P(H_i)}{\sum P(X=\hat{\theta}|H_j)P(H_j)} \qquad \text{(Equation 1)}$$

where $X$ is a random variable varying over all $n$ possible sample estimates of $\theta$, the sum is taken over all the hypotheses, $H_j$, and $P(H_j)$ is the prior probability of the corresponding hypothesis. Assumption 4 means that we can cancel out the (equal) prior probabilities, so this reduces to

$$P\left(H_i \mid X = \hat{\theta}\right) = \frac{P(X=\hat{\theta}|H_i)}{\sum P(X=\hat{\theta}|H_j)} \qquad \text{(Equation 2)}$$

Using Assumption 3, the distribution under $H_i$ is simply the distribution under $H_0$ moved $i$ units to the right so we can write the numerator of the right hand side as *P* values, confidence intervals, or confidence levels for hypotheses?

$$P\big(X = \hat{\theta}\,\big|H_i\big) = P\big(X = \hat{\theta} - i\,\big|H_0\big) \qquad \text{(Equation 3)}$$

(For example, if *i* = 0.5 in Figure 1, the distribution will be moved 0.5 units to the right so the left hand side of Equation 3 will be the probability corresponding to 1.1 − 0.5 = 0.6 in the null hypothesis distribution.) Also, because the distribution is symmetrical (Assumption 2):

$$P\big(X = \hat{\theta}\,\big|H_i\big) = P\big(X = i - \hat{\theta}\,\big|H_0\big) \qquad \text{(Equation 4)}$$

This means that Equation 2 becomes

$$P\big(H_i \mid X = \hat{\theta}\big) = \frac{P\big(X = i - \hat{\theta}\,\big|H_0\big)}{\sum P\big(X = \hat{\theta}\,\big|H_j\big)} \qquad \text{(Equation 5)}$$

If we sum this equation for all values of *i*, the left hand side obviously sums to one, as does the numerator on the right hand side, which means that the denominator on the right hand side also sums to one, and the equation reduces to

$$P\big(H_i\big|X = \hat{\theta}\big) = P\big(X = i - \hat{\theta}\,\big|H_0\big) \qquad \text{(Equation 6)}$$

This equation tells us that the posterior probability of each hypothesis is given by a discrete version of Figure 1 shifted $\hat{\theta}$ units to the right, which is a discrete version of Figure 2. As the above argument does not depend on whether we measure to the nearest tenth, or hundredth or thousandth, we can get as close as we like to a continuous distribution, so we can view the continuous version, Figure 2, as the posterior distribution, or, using a Bayesian interpretation of confidence, as a confidence distribution.

In practice, Assumptions 1-3 are satisfied when the normal or *t* distributions are used to model the null hypothesis. There may, of course, be other sets of assumptions which lead to the same conclusion (e.g. Bolstad, 2007: 227), but Assumptions 1-3 have the advantage that they have a simple graphical interpretation.

# Notes

---

[1]   This example is based on Table 3 of Eggins et al (2008), although I have changed the details of the experiment to give an illustration which can be appreciated by readers who have not read this article.

[2]   This conception of NHSTs is usually traced back to Fisher, and seems to be the standard in most social sciences, although it is sometimes combined with the "Neyman-Pearson" position and errors of Types I and II in a manner which is not entirely consistent – see Cortina and Folger (1998: 340-341).

[3]   The value of *t* in the paper from which the example is drawn (Eggins et al, 2008) is 2.40, and the difference of the means is 1.07, which means that the standard error of the difference is 0.446: this allows us to use the *t* distribution to calculate these statistics, and others cited below.

*P* values, confidence intervals, or confidence levels for hypotheses?

---

[4]    The first, third and fourth of these are estimated using the mean of results from the two CI limits given in the article.

[5]    The Excel formula for the *p* value is =1-BINOMDIST(6,20,0.1,TRUE). Figure 3 is based on the binomial distribution.